\documentstyle{article}
\input tcilatex

\begin{document}

\baselineskip=23pt

\begin{flushleft}
{\bf {\Huge Evidence consistent with the cosmological interpretation of
quasar redshifts}}\\
\vspace{4mm} 
{\bf Yi-Ping Qin$^{1,2,3}$, Guang-Zhong Xie$^{1,2}$, Xue-Tang Zheng$^{4}$,
and En-Wei Liang$^{1,2,5}$}

{\bf $^{1}$ Yunnan Observatory, Chinese Academy of Sciences, Kunming, Yunnan
650011, P. R. China}

{\bf $^{2}$ National Astronomical Observatories, Chinese Academy of Sciences 
}

{\bf $^{3}$ Chinese Academy of Science-Peking University joint Beijing
Astrophysical Center }

{\bf $^{4}$Department of Physics, Nanjing University of Science and
Technology, Nanjing, Jiangsu 210014, P. R. China}

{\bf $^{5}$Physics Department, Guangxi University, Nanning, Guangxi
530004, P. R. China}
\\

\vspace{10mm} 

{\bf {\Large ABSTRACT}}

\end{flushleft}

In this letter, the old issue of whether redshifts of quasars are of
cosmological origin is investigated. We make a plot of absorption redshifts
versus emission redshifts for quasars with large amounts of data. Our study
shows that, almost all absorption redshifts are smaller than the
corresponding emission redshifts. The relation between the absorption and
emission redshifts predicted by current cosmological models is well obeyed.
The result confirms that redshifts of quasars are indeed distance
indicators. It might be the most obvious evidence found so far to be
consistent with the cosmological interpretation of quasar redshifts.

\begin{flushleft}


\end{flushleft}

\vspace{10mm}

\section{INTRODUCTION}

Redshifts of quasars are commonly believed to be of cosmological origin.
They may be a natural observational phenomenon of distant objects predicted
by current cosmological models (e.g., Weinberg 1972). Debate of this issue
has continued for a long time. Fresh negative evidence keeps coming in,
throwing doubt on this interpretation (e.g., Arp 1968b, 1988; Narlikar 1986;
Arp et al. 1990; Duari et al. 1992; Chu et al. 1998). Perhaps the strongest
challenge comes from the evidence of the association between quasars or BL
Lac objects and galaxies, where one often finds high-redshifts for the
former and low-redshifts for the latter (see, e.g., previously, Arp 1966,
1967, 1968a; and recently, Arp 1997, 1999a, 1999b; Burbidge 1997; Radecke
1997). At the same time, more and more positive evidence continues to
emerge. One can find most evidence from the studies of quasars, or from the
investigations of cosmological models, or from the searches for
gravitational lensing of distant objects, where any self-consistent results
can be taken as indirect positive evidence supporting current cosmological
models, while some of non-self-consistent results might be considered as
indirect negative evidence. Since most of the studies provided
self-consistent results rather than non-self-consistent ones (for recent
reviews see, e.g., Trimble and McFadden 1998; Trimble and Aschwanden 1999),
we believe that the positive evidence must be an overwhelming majority. The
most convincing evidence for the cosmological nature of redshifts for
extragalactic sources may probably be the discovery of type Ia supernovae as
standard cosmological candles (see, e.g., Garnavich et al. 1998; Riess et
al. 1998a, 1998b; Perlmutter et al. 1999). A direct study of this issue is
by the well-known means of the Hubble diagram of quasars. Recently, with a
new approach in the study of the diagram, Qin et al. (1997) found that the
relation between the redshift and the observed brightness of quasars is
consistent with the cosmological hypothesis. In the following we will study
the relation between absorption and emission redshifts of quasars. The
relation might probably provide more direct evidence demonstrating the
nature of the redshifts.

\section{RELATION BETWEEN ABSORPTION AND EMISSION REDSHIFTS}

A natural observational effect of distant objects is that their continuum
spectrum can be absorbed by the foreground medium. If the redshifts are
cosmological distance indicators (the larger the distance, the greater the
redshift), then all the absorption redshifts, $z_{abs}$, of a source should
be smaller than its emission redshift, $z_{em}$. If they are not distance
indicators, then there should be no such obvious relation. An investigation
of the plot of $z_{em}-z_{abs}$ for a large number of distant sources should
give a direct and definite answer to this issue.

In Table 1 of Hewitt and Burbidge (1993), there are 401 sources with both
absorption and emission redshifts available. A given quasar can only have
one emission redshift and may have several absorption redshifts. This is a
natural consequence of the absorbers being foreground objects, whether or
not redshifts are distance indicators. From the above sample of 401 sources
we must discard 0820+225, because it is a member of the well-known 1Jy BL
Lacertae sample (see Stickel et al. 1991). For 2359-022, the $z_{em}$ given
in the catalog must be wrong (see Wolfe et al. 1993) and we will use the
value given by Wolfe et al. (1993), $z_{em}=2.81$, and $z_{abs}=2.154$, $%
2.095$. Our final sample contains a total of 1,306 absorption redshifts
belonging to 400 sources.

In this sample, there are 66 sources (16.5\% of the total) having one or
more of their absorption redshifts larger than the emission redshifts. We
call such sources as ``anomalous sources''. The total number of such
absorption redshifts (which we call ``anomalous absorption redshifts'') is
86, or 6.6\% of all the absorption redshifts. A plot of $z_{abs}$ versus $%
z_{em}$ for this sample is displayed in Figure 1.

As expected, the majority (in fact 93.4\%) of the data points fall below the 
$z_{em}=z_{abs}$ line, i.e., they have $z_{em}<z_{abs}$. This is precisely
the relation between the absorption and emission redshifts predicted by
current cosmological models.

\section{DISCUSSION AND CONCLUSIONS}

In last section, we investigate whether the redshifts of quasars are of
cosmological origin by making a plot of absorption redshifts versus emission
redshifts for the objects with large amounts of data. The study shows that,
almost all absorption redshifts are smaller than the corresponding emission
redshifts. The relation between the absorption and emission redshifts
predicted by current cosmological models is well obeyed.

Of course, if an absorber is located very close to the quasar and has a very
large relative motion towards the quasar, then the absorption redshift may
become larger than the emission redshift. But such circumstances would seem
to be too rare to account for the appreciable fraction (6.6\%) of anomalous
absorption redshifts. Other factors must be at work, and we suspect that the
scarcity of data points in the bottom right portion of the plot may be just
such a factor.

We also find from the figure that the anomalous absorption redshifts are
quite close to the corresponding emission redshifts in value. Let $\triangle
z\equiv z_{abs}-z_{em}$ and $\triangle z/z\equiv (z_{abs}-z_{em})/z_{em}$.
For the 86 anomalous absorption redshifts, we find: $\left\langle \triangle
z\right\rangle =0.0137$, $(\triangle z)_{\max }=0.1450$, $\left\langle
\triangle z/z\right\rangle =0.0061$, and $(\triangle z/z)_{\max }=0.0661$.
This shows that anomalous absorption redshifts only differ from their
corresponding emission redshifts by minute amounts. Redshifts, including
anomalous absorption redshifts, are mainly due to cosmological distances
rather than relative motions.

Our sample is not statistically complete and many selection effects must be
at work. Selection effects might cause incompleteness in the magnitude
distribution or the redshift distribution of the sample. The former would
not affect the relation between the absorption and emission redshifts. As
regards the latter, any selection effects affecting the observation of
emission redshifts should, in our opinion, equally affect the observation of
absorption redshifts. So, there are no reasons to think that our plot is
seriously vitiated by selection effects. Effects such as a possible density
evolution of quasars or the Malmquist bias might probably alter the values
of the above two percentages, but since they do not have any bearing on the
relation between the two redshifts, they would not change the basic fact
that emission redshifts of quasars are larger than the great majority of
their corresponding absorption redshifts.

The above result indicates that redshifts of quasars are distance
indicators, which is consistent with current cosmological models. However,
this is also true for the tired-light hypothesis. The plot made in this
letter can not tell the difference between the two hypotheses. To exclude
the tired-light hypothesis, one might rely on other evidence, which is out
of the scope of this letter.

One may ask why should we exclude the possibility that the absorbing
material is ejected from the quasar, and that we are measuring a Doppler
redshift. If emission redshifts are distance indicators and absorption
redshifts are caused by the ejected material, then a similar result would be
expected, but the precondition itself (emission redshifts are distance
indicators) does not contradict with the distance indicator scenario
described by cosmological models or the tired-light hypothesis. If emission
redshifts are not distance indicators and absorption redshifts are caused by
the ejected material, then the universe would not be expanding and also
there would be no tired-light happened. In this way, the universe must be
either stationary or shrinking, then the adsorption redshifts should be
negative (relative to blue-shift) rather than positive, but this is not true.

The fact that emission redshifts of quasars are larger than the great
majority of their corresponding absorption redshifts obviously contradicts
with the work of Arp et al. (see Section 1). If there indeed exists
association between high-redshift quasars and low-redshift galaxies, then
the absorption redshifts of the quasars would be close to the emission
redshifts of the galaxies or even less than them (when the absorbing
material is ejected from the galaxies).

We then come to the conclusion that redshifts of quasars are definitely
distance indicators. The result shown in our plot might probably be the most
obvious and direct evidence found so far supporting the distance indicator
hypothesis, which is consistent with current cosmological models.

\vspace{20mm}

\begin{flushleft}
{\bf {\Large ACKNOWLEDGMENTS}}\\
\end{flushleft}

It is our pleasure to thank Dr. Tao Kiang for helpful discussion. This work
was supported by the United Laboratory of Optical Astronomy, CAS, the
Natural Science Foundation of China, and the Natural Science Foundation of
Yunnan.

\vspace{60mm} 
\begin{flushleft}
{\bf {\Large FIGURE CAPTION}}\\
\end{flushleft}

\begin{verse}
{\bf Figure 1.} The $z_{abs}-z_{em}$ plot for 400 quasars. The solid line is
the $z_{abs}=z_{em}$ line.\\ 
\end{verse}

\newpage

\begin{flushleft}
{\bf {\Large REFERENCES}}\\
\end{flushleft}

\begin{verse}
Arp, H. 1966, Science, 151, 1214\\

Arp, H. 1967, ApJ, 148, 321\\

Arp, H. 1968a, ApJ, 152, 633\\

Arp, H. 1968b, ApJ, 153, L33\\

Arp, H. 1988, Quasars, Redshifts, and Controversies, Cambridge University
Press, Cambridge\\

Arp, H. 1997, A\&A, 319, 33\\

Arp, H. 1999a, A\&A, 341, L5\\

Arp, H. 1999b, ApJ, 525, 594\\

Arp, H. et al. 1990, A\&A, 239, 33\\

Burbidge, E. M. 1997, ApJ, 484, L99\\

Chu, Y. et al. 1998, ApJ, 500, 596\\

Duari, D., Gupla, P. D., Narlikar, J. V. 1992, ApJ, 384, 35\\

Garnavich, P. et al. 1998, ApJ, 493, 53\\

Hewitt, A., Burbidge, G. 1993, ApJS, 87, 451\\

Narlikar, J. V. 1986, in Swarup, G., Kapahi, V. K., eds, IAU Symp. 119,
Quasars. Reidel, Dordrecht, p.463\\

Perlmutter, S. et al. 1999, ApJ, 517, 565\\

Qin, Y. P., Xie, G. Z., Wang, J. C. et al. 1997, ApSS, 253, 19\\

Radecke, H.-D. 1997, A\&A, 319, 18\\

Riess, A. G. et al. 1998a, ApJ, 504, 935\\

Riess, A. G. et al. 1998b, AJ, 116, 1009\\

Stickel, M., Padovani, P., Urry, C. M., Fried, J. W., Kuhr, H. 1991, ApJ,
374, 431\\

Trimble, V., Aschwanden, M. 1999, PASP, 111, 385\\

Trimble, V., McFadden, L.-A. 1998, PASP, 110, 223\\

Weinberg, S. 1972, Gravitation and Cosmology: Principles and Applications of
the General Theory of Relativity. John Wiley, New York\\

Wolfe, A. M., Turnshek, D. A., Lanzetta, K. M., Lu, L. 1993, ApJ, 404, 480\\
\end{verse}

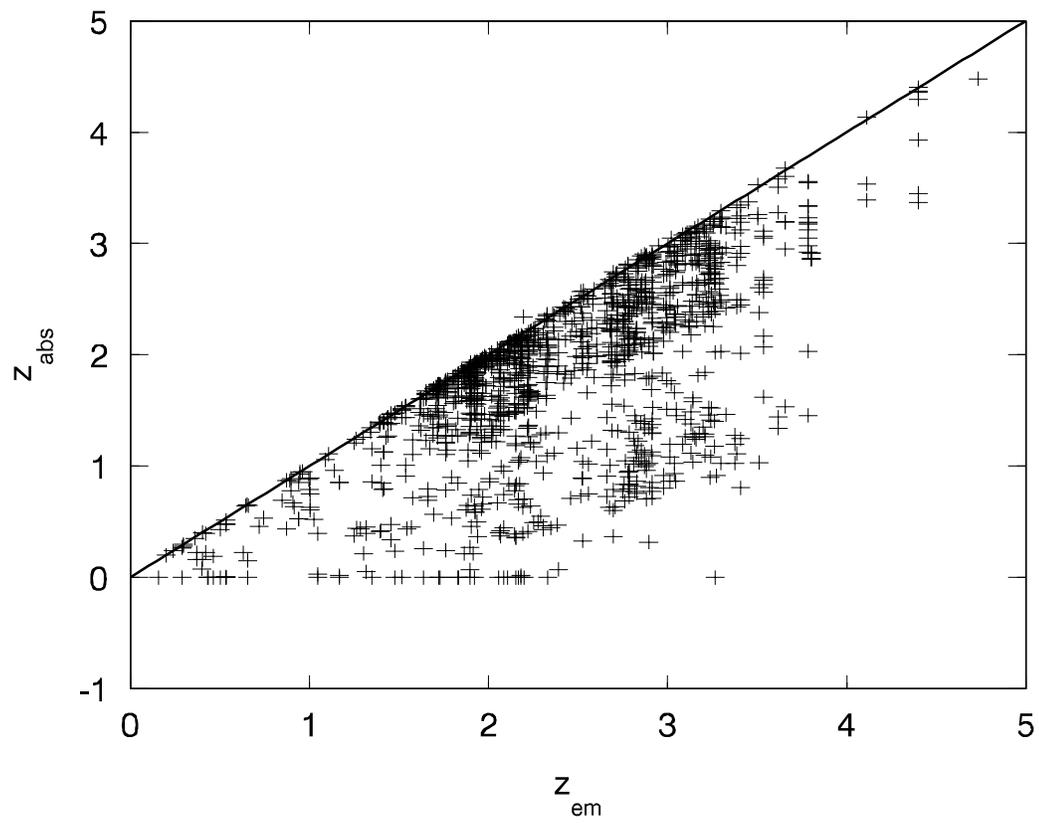
\end{document}